\documentclass[12pt]{article}
\usepackage[utf8]{inputenc}
\usepackage{textcomp}
\usepackage{amsmath}
\usepackage{amssymb}
\usepackage{graphicx}
\usepackage{braket}
\usepackage{url} 
\usepackage{comment} 
\usepackage{subfigure}
\usepackage[makeroom]{cancel}
\usepackage{setspace} 
\usepackage{hyperref}
\usepackage{mathptmx}
 
\usepackage{authblk}

\usepackage[legalpaper, portrait, margin=1in]{geometry}

\usepackage[style=authoryear,maxbibnames=200]{biblatex}
\usepackage[bblfile=main]{biblatex-readbbl}

\begin{document}

\pagenumbering{roman}

%\title{\huge \vspace{-2.0cm} From statistical dependence to the space of possible superdeterministic theories}

\title{\huge \vspace{-2.0cm} 
From statistical dependence to the space of possible superdeterministic theories\thanks{This paper is forthcoming in \href{https://link.springer.com/journal/13194}{\textit{The European Journal for Philosophy of Science}}.}}

%\author{}

\author[1]{Mordecai Waegell}
\author[1,2]{Kelvin J. McQueen}

%\affil[1]{Institute for Quantum Studies, Chapman University, Orange, CA 92866, USA}
%\affil[2]{Philosophy Department, Chapman University, Orange, CA 92866, USA}

\affil[1]{Institute for Quantum Studies, Chapman University}
\affil[2]{Philosophy Department, Chapman University}

%Shunk affiliation text and removed irrelevant address info, thereby creating a beautiful pyriamid shape (KM 9/1)

\date{\today}

\maketitle

\begin{spacing}{1.2}
    \begin{abstract}
Bell's theorem demonstrates that any physical theory that is consistent with the predictions of quantum mechanics, and which satisfies some apparently innocuous assumptions, must violate the principle of \textit{local causality}. It may therefore be possible to maintain local causality by rejecting one of these other assumptions instead. One possibility that has recently received significant attention involves rejecting the principle of \textit{statistical independence} (SI). In this paper, we consider the \textit{frequency interpretation} of SI, which states that $\rho(\lambda) \approx \rho(\lambda | Z)$, where $\rho(\lambda)$ is the relative frequency of an element of an ensemble being in the state $\lambda$, and $Z$ is a label that separates the ensemble into apparently randomly selected sub-ensembles. SI is violated when the sub-ensemble frequency $\rho(\lambda | Z)$ fails to be representative of the ensemble frequency $\rho(\lambda)$. We argue that physical theories that systematically violate SI should all be understood as \textit{superdeterministic}. This perspective on SI sheds light on a number of issues that are being debated in the superdeterminism literature, especially concerning its scope and philosophical consequences. Regarding scope, we argue that superdeterministic theories fall into three categories, deterministic theories with fine-tuned initial conditions, fluke theories, and nomic exclusion theories. We also argue that retrocausal and invariant set theories need not violate SI, which is contrary to how they are normally presented. Regarding philosophical implications, we argue that superdeterminism is incompatible with free will according to some prominent compatibilist accounts. We also argue that although superdeterminism is conspiratorial, it is not unscientific, but pre-scientific.
\end{abstract}
\end{spacing}

    {\bf Keywords:} Superdeterminism, statistical independence, measurement independence, Bell's theorem, locality, free will, retrocausality.

\pagebreak

%Contents
\begin{spacing}{1.8}
\tableofcontents
\end{spacing}
\newpage

\section{Introduction}\label{intro}
\pagenumbering{arabic}
\setcounter{page}{1}

Bell's theorem (\cite{bell1964einstein}, \cite{Bell_1971}) shows that any physical theory that is consistent with the predictions of quantum mechanics, and which satisfies some apparently innocuous assumptions, must violate the principle of \textit{local causality}.

Some argue that these other assumptions are \textit{so} innocuous that they cannot coherently be denied, so that we may confidently assert that Bell's theorem has \textit{established} that physics is nonlocal (\textcite{maudlin2014bell}). Others think that local causality is worth preserving, leading them to question these other assumptions, and to inquire into the feasibility of a local quantum theory that denies one of them. The apparently innocuous assumption that is the focus of this paper is the principle of \textit{statistical independence} (or SI for short).\footnote{Other options for preserving locality include denying the assumption that there is only one world (\cite{Vaidman2016}, \cite{WAEGELL2020}) and denying that there is no retrocausality (see section \ref{retrocausal} below).} 

There has been much debate over how to best understand SI. We think SI is best understood as stating that \textit{random sampling procedures yield representative samples}, and that viewing it this way helps to clarify its consequences. This is a familiar idea in science. It means that we can collect up entities of some type, so that experiments on those entities can potentially reveal general facts about \textit{all} entities of that type. It is the basis of inductive generalization. It is no wonder that some have thought that denying SI would ``undercut the whole scientific method" (\cite{maudlin2014bell}). More recently, several authors have tried to make this objection more explicit (\cite{baas2023does}, \cite{allori2024hidden}). 

Others disagree, and believe that so-called \textit{superdeterministic theories} can be rigorously developed. These are theories that deny SI (instead of local causality), and which aim to explain the outcomes of Bell experiments.\footnote{Attempts at developing superdeterministic theories can be found in \textcite{t2016cellular}, \textcite{donadi2020superdeterministic}, 
Palmer (\citeyear{palmer2022discretised}, \citeyear{palmer2024superdeterminism}), and \textcite{ciepielewski2023}.} Some have also offered explicit responses to the above-mentioned criticisms of superdeterminism (\cite{hossenfelder2020rethinking},
\cite{andreoletti2022superdeterminism}, \cite{nikolaev2023aspects}).

In light of these developments a number of debates have arisen about superdeterministic theories, what they do and do not require, whether they allow free will and whether they are conspiratorial. Our aim in this paper is to make clear what it really means for a theory to violate SI, i.e., to imply that random sampling procedures fail to give representative samples (especially in the context of Bell experiments). In doing so, we aim to contribute to all of the above mentioned issues.

In our view, Bell coined the term `superdeterministic' to refer specifically to the idea that nature may not provide sufficient sources of randomness, such that random selection processes in experiments could never be relied upon to give representative samples. This is the exact scope of SI, so it seems natural to equate SI violation with superdeterminism as we do here.  We think this passage from \textcite{bell1990nouvelle} makes this clear:
\begin{quote}
    ``An essential element in the reasoning here is that [the measurement settings] are free variables. One can envisage then theories in which there just are no free variables [...] In such ‘superdeterministic’ theories the apparent free will of experimenters, and any other apparent randomness, would be illusory. Perhaps such a theory could be both locally causal and in agreement with quantum mechanical predictions. However I do not expect to see a serious theory of this kind. I would expect a serious theory to permit ‘deterministic chaos’ or ‘pseudorandomness’, for complicated subsystems (e.g. computers) which would provide variables sufficiently free for the purpose at hand. But I do not have a theorem about that.''
\end{quote}
Bell is particularly concerned here about getting a sufficient type of randomness out of deterministic theories so that they are not also superdeterministic.  The frequency interpretation of SI addresses this concern, since given a sub-ensemble along with the ensemble from which it was drawn, one can evaluate SI, and it does not matter whether those ensembles were selected using a deterministic or indeterministic process. The success or failure of a random selection process to give a representative sample is independent of whether or not nature is deterministic.

In section \ref{SI2SD} we build on existing work that aims to show that SI is not a probabilistic claim, but is instead a claim about whether or not our samples are representative of the populations we have sampled from (\cite{chen2021bell}, \cite{ciepielewski2023}, \cite{allori2024hidden}). But we do so in a way that enables us to challenge the widespread claim that superdeterminism is restricted to deterministic theories with fine-tuned initial conditions. Section \ref{initialconditions} explains theories of this sort and responds to arguments that superdeterminism must take this form. Section \ref{flukes} explains a quite different class of superdeterministic theories based on statistical flukes. Section \ref{Goblins} explains a third category of superdeterminism, which we call ``nomic exclusion" theories.

Section \ref{mistakenidentity} then takes two different types of theories that are often said to violate SI (retrocausal theories and invariant set theory), and argues that this need not be the case, and that neither type of theory is necessarily superdeterministic.

Section \ref{consequences} then moves to philosophical consequences of superdeterminism. In section \ref{freewill} we challenge the widely held claim that superdeterminism has no bearing on free will. In particular, we argue that some prominent compatibilist accounts of free will entail that superdeterminism restricts our free will. Section \ref{conspiracy} argues that superdeterministic theories all have a conspiratorial element to them, and we respond to arguments to the contrary (\cite{andreoletti2022superdeterminism}). Finally, section \ref{scientific} addresses the scientific status of superdeterminism. We argue that although superdeterminism does not necessarily undermine the scientific method, that does not make it scientific. We think it is best understood to be in a pre-scientific state.

\section{From statistical dependence to superdeterminism}\label{SI2SD}

\subsection{What is statistical independence?}\label{SI}

The assumption of statistical independence (SI) was first made explicit by \textcite[note 13]{Clauser_Horne_1974}. \textcite{Bell_1977} explicitly acknowledges the need for it, but describes it as ``a point of leverage for ‘free willed experimenters’". As a consequence, SI has sometimes been called \textit{the Free Will assumption} or the \textit{the Freedom of Choice Assumption} (\cite{sep-bell-theorem}). Recent research has almost unanimously rejected Bell's characterization, arguing that free will has nothing to do with it, a point we return to in section \ref{freewill}.

SI is now standardly defined as $\rho(\lambda|X,Y) = \rho(\lambda)$, where $X$ and $Y$ are measurement settings and $\lambda$ are what determine the measurable properties of the measured systems. However, there have been different interpretations of $\rho$: is it a probability or is it a relative frequency? And if it is a probability, is it an epistemic or an ontic probability? 

The probabilistic interpretation has a long history. For example, as \textcite{chen2021bell} explains, many sought to maintain \textit{both} local causality and SI, by rejecting implicit assumptions about classical probability theory that Bell may have made in his proof. Chen responds by proving Bell's theorem without using probability theory. In particular, Chen formulates each assumption of Bell's theorem without using probabilities, and in the case of SI, he adopts the frequency interpretation. The probabilistic interpretation has also been criticized by \textcite[p444]{ciepielewski2023}, who instead defend the frequency interpretation. The frequency interpretation has also recently been adopted and defended by \textcite{allori2024hidden}).\footnote{An early defense of the frequency approach is given by Maudlin in an illuminating online debate with Tim Palmer who, as we shall see in section \ref{Palmer}, instead adopts the ontic probability approach. See \href{youtube.com/watch?v=883R3JlZHXE}{youtube.com/watch?v=883R3JlZHXE}.} We find that this interpretation of $\rho$ provides a fresh perspective on many topics discussed in the context of superdeterminism and our goal in this paper is to clarify the interpretation and its consequences for these topics.

On the relative frequency interpretation, SI is easy to interpret, and applies well beyond quantum mechanics. Simplified, it states that $\rho(\lambda) \approx \rho(\lambda | Z)$, where $\rho(\lambda)$ is the relative frequency that an element of an ensemble is in the state $\lambda$, and $Z$ is a label that separates the ensemble into apparently randomly selected sub-ensembles.

For example, in an ensemble of apples, $\lambda$ might be `red apple', and $Z$ might be `left box'. SI then says that if the sub-ensembles are sufficiently large, and selected in an apparently random way, then the relative frequency of $\lambda$ in every sub-ensemble should be approximately the same as its relative frequency in the entire ensemble. So, if half the apples are red, then approximately half the apples in the left box are red. This is really just another way of saying that the random sampling procedure marked by \textit{Z} generates a \textit{representative sample} of a larger ensemble. Consequently, a theory which systematically violates SI, that is, a superdeterministic theory, 
must entail that the random sampling procedure fails to yield representative samples.

%This interpretation of SI may seem rather anthropocentric, which may lead some to prefer the ontic interpretation, which is directly defined in terms of a theory's ontic probabilities. But even the ontic interpretation is defined in terms of the probability of $\lambda$ \textit{given the choice of certain settings}. So an element of anthropocentrism seems unavoidable here. What's important is whether SI is formulated as a yes or no question for a physical theory. And in principle, we can always check, for any given physical theory, whether the mechanisms it postulates entail that the relevant random sampling procedures fail to yield representative samples. So, the frequency definition of SI is sufficiently objective. 

 Random sampling procedures are a familiar practice in science. They begin by defining the total ensemble or population of interest, which we want to learn general facts about. Typically, the population of interest is too large for us to examine every member. And so we instead only examine a sample of the population. If the sample is representative of the population, then we can be confident that what we discover about our sample represents a discovery about the total population. Acquiring a \textit{random} sample is a way to help ensure our sample is representative of the population. As \textcite{chen2021bell} and \textcite{allori2024hidden} emphasize, random sampling is what allows scientists to use induction.

How do we ensure that our samples are randomly selected from the population of interest? Consider a simple yet notorious example. In 1936, the magazine \textit{Literary Digest} obtained a sizable sample of polling data of the voting population of America: 2.4 million voters. In this sample, 57\% would vote for the Republican, Alf Landon, so the magazine predicted a loss for the Democrat, Franklin Roosevelt. Yet Roosevelt won 62\% to 37\%. This is a case in which random sampling \textit{failed to take place for the intended population}. The sampling method that was used meant that effectively everyone in the sample was wealthy. Since wealth is a relevant factor for voting preference, random sampling over the US population did not take place. Instead, random sampling only happened (if at all) over the \textit{wealthy} U.S. population. This is an important distinction. For as we understand superdeterminism, representative samples are not obtained \textit{even when random sampling over the relevant population is apparently successful}, e.g., even when the sample includes a proportional representation of wealthy and non-wealthy people (and all other observable properties). In other words, if superdeterminism is true, there isn't anything we can do to obtain a representative sample (for the relevant domain of interest).

SI is relevant to Bell experiments because to conclude that nature is nonlocal, we assume that the choice of measurement settings is independent of the state of the system to be measured.  If our measurement settings are the angles at which we position our polarizers, and our measurement outcomes are whether or not entangled photons are absorbed by those polarizers, then it is essential that the settings are chosen independently of the properties of the photons used in the experiment. This independence prevents any systematic bias that could lead to unrepresentative samples—where, for instance, photons sent to a polarizer at one angle are fundamentally different from those sent to a polarizer at another angle. To achieve this, settings must be selected randomly. Various methods are employed to ensure randomness in Bell experiments: some use pseudo-random number generators, which rely on algorithms to produce sequences of bits that are statistically independent; others employ quantum processes involving inherently unpredictable outcomes according to quantum theory. More innovative approaches include using cosmic photons, such as signals from distant stars or quasars, where the vast distances and time intervals make any causal link between these photons and the experiment highly implausible. These strategies are designed to increase our confidence that the settings are chosen randomly and independently of the photon state, and thus that SI is obeyed for these Bell experiments.

When we perform the above-mentioned experiments on certain sorts of entangled photon pairs, we see certain very specific results. For example, if the two polarizers point in the same direction (modulo GR corrections), then the measurement outcomes on the two photons will be the same: either both will be absorbed or both will be transmitted (\cite{chen2021bell}). When Bell's theorem is proved in the context of this type of experiment, it must be assumed that these results are indicative of all photons, and not just the ones that happened to be measured in this way. Superdeterministic theories deny this assumption. 

It is widely believed that superdeterminism must be deterministic and must involve fine-tuned or atypical initial conditions. For example, \textcite{baas2023does} assert that ``superdeterminism is the conjunction of determinism and the atypicality of cosmological initial conditions". Meanwhile, \textcite[p156]{allori2024hidden} asserts that ``superdeterminism can succeed only [by] resorting to fine-tuned initial conditions". (Allori's claim is based on a formal result by \textcite{Durr_Teufel_2009}, which we discuss in the next section.) 

But this is an unnecessary restriction that can lead one astray when making general claims about superdeterminism, as we shall see. Superdeterminism was introduced by Bell as a loophole to Bell's theorem, one that allows us to maintain locality by rejecting SI.  Fine-tuned initial conditions may be one way of reaching this result, but it is not the only way. 

We have argued that the key defining feature of superdeterminism is the violation of SI.\footnote{Several authors have put forth their own proposed definitions of superdeterminism e.g. \textcite{wiseman2017causarum}, \textcite{WAEGELL2020}, \textcite{sen2020superdeterministic},  \textcite{wharton2020colloquium}, \textcite{adlam2024taxonomy}, which are each distinct in some way.} We think there are two ways other than fine-tuned initial conditions that can yield this result. We therefore propose the following:

\begin{quote}
A theory is superdeterministic if and only if it entails that random sampling procedures systematically fail to yield representative samples, in virtue of either 

1) Fine-tuned initial conditions in a deterministic theory; or

2) Statistical flukes (in a deterministic or indeterministic theory); or 

3) Nomic exclusion (in a deterministic or indeterministic theory).
\end{quote}

In the next section (\ref{3forms}), we consider each of these in turn. Then in the following section \ref{mistakenidentity} we consider two cases that are difficult to categorize, retrocausality and invariant set theory.

\subsection{Three forms of superdeterminism}\label{3forms}

\subsubsection{Determinism with fine-tuned initial conditions}\label{initialconditions}

Superdeterminism is often taken to entail fine-tuned or atypical initial conditions in a deterministic universe. Take the fact that for the relevant entangled photon pairs, we always see that if the two polarizers point in the same direction, then either both photons will be absorbed or both will be transmitted. Perhaps only some photons behave this way, while most don't, it's just that the ones that don't never make their way into this type of experimental set up, due to the atypical initial conditions of the universe.

Is it inevitable that superdeterminism be a deterministic, local theory with fine-tuned atypical initial conditions? While many have taken these to be defining features of superdeterminism, we think none of them are necessary and in the next two sections we present counterexamples. But first we should consider a result in \textcite{Durr_Teufel_2009}, which \textcite{allori2024hidden} has described as ``a formal result establishing that SI holds for typical initial conditions, so that superdeterminism can succeed only [by] resorting to fine-tuned initial conditions".

Dürr and Teufel (2009) demonstrate a result for deterministic frameworks, using the example of a Galton board—a device where balls bounce off pegs and eventually settle into boxes at the bottom, forming a distribution that matches probabilistic predictions. In their analysis, they trace the randomness of the balls' paths back to the initial conditions of the balls entering the board and further to the cosmological initial conditions of the universe. By introducing a typicality measure for these cosmological initial conditions, they show that typical initial conditions will lead to a ball bouncing left or right off the pegs it encounters a roughly equal number of times, whereas atypical initial conditions could result in a ball bouncing left off of every peg in the board. Dürr and Teufel do not explicitly connect this result to SI. But perhaps Allori's thought is that if we were to see the balls always bouncing left, then we would have a sample that fails to represent typical initial conditions, in virtue of atypical fine-tuned initial conditions.

However, even if Allori's interpretation of Dürr and Teufel's result were correct, it still falls short of showing that superdeterminism (the denial of SI) entails atypical or fine-tuned initial conditions. First, the result assumes determinism, but in section \ref{flukes} we show how indeterminism without fine-tuned initial conditions can violate SI. Second, even assuming determinism, it assumes there are no special systems (``goblins") whose behavior leads to SI violations, without atypical initial conditions. This case is explained in section \ref{Goblins}. 

In the remainder of this section, we consider a recent specific superdeterministic model, based on fine-tuned initial conditions (\cite{ciepielewski2023}). Part of the interest of this model is that it postulates atypical initial conditions that are not very complicated. It therefore offers a response to a common criticism of superdeterminism that it requires overly complex initial conditions (\cite{chen2021bell}). Following \textcite{chen2021bell}, we will refer to it as Leibnizian Quantum Mechanics (LQM), due to some parallels to Leibniz’s Monadology (\cite{leibniz1714monadology}).  LQM is a superdeterministic model that replicates quantum predictions using a unique structure where, at each point in physical space, there exists an internal space (a copy or simulation of a universe). In LQM, the evolution of the universe in physical space is determined by the dynamics within these internal spaces, which each follow Bohmian mechanics. Essentially, LQM replaces the single universal wave function in Bohmian mechanics with a continuous infinity of wave functions, each belonging to an internal space and evolving independently within that space following the Schr\"{o}dinger equation. There are no particles moving in physical space, and instead the configuration of mass density at a given location in physical space is given by the internal state at the corresponding location. The initial conditions are atypical in that they are identical for every internal space and there is no reason to expect this, given that they are all independent. This model shows that the atypical initial conditions of a superdeterministic theory need not be overly complex, although as \textcite{chen2021bell} notes, an excessive complexity lies instead in the Leibnizian ontology. 

How exactly is this model superdeterministic?  According to the authors, ``To see that SI is indeed violated once homogeneous
conditions are provided, we note that the state $\lambda$ of the particles involves the
specification of both $\Phi$ (the universal wavefunction) and $Z$ (the Bohmian position variables) in the region where the particles are created. But the
homogeneity of initial conditions implies that such $\Phi$ and $Z$ determine such fields
everywhere, including in the regions where $a$ and $b$ (settings) are located. We conclude that
in this case, $\lambda$, $a$, and $b$ cannot be independent" (\cite[p454]{ciepielewski2023}). Bohmian mechanics, as typically understood, does not violate SI. This is because the wavefunction of the Bell state ($\lambda$) and the experimental settings ($a$ and $b$) are (effectively) separable from one another and can be independently varied (\cite{esfeld2015bell}, \cite{allori2024hidden}). LQM, on the other hand, apparently does violate SI, despite the similar ontology. This is because the $\lambda$ that is associated with a particle corresponds to the $\Phi$ and $Z$ in the region of the particle, and these correspond to an \textit{entire internal Bohmian universe}. Given homogeneous initial conditions, every point in physical space has the same internal state $\Phi$ and hidden variable $Z$, so the $\lambda$ of any given particle encodes information about all settings, no matter where in physical space they are chosen.  Thus, given the atypical homogeneity condition, neither the settings nor $\lambda$ can be varied, no matter where they are in physical space, so they remain perfectly correlated.  For a typical inhomogeneous initial condition, no such correlations would exist, since the wavefunction at one location generally encodes no information about any other location - and it seems physical space is generally an abstract mess, which contains nothing resembling a Bell experiment.

However, the authors adopt the frequency interpretation of SI, in which $\rho(\lambda)$ represents ``the actual distribution of states over the measured ensemble" (p444). But in the explanation of the violation of SI quoted above, $\lambda$ is understood in terms of $\Phi$ and $Z$, i.e., entire internal Bohmian universes, which given homogeneous initial conditions, cannot be different for different particles. So there is a problem here in explaining how LQM violates SI under their preferred frequency interpretation. Part of the difficulty here is the very exotic ontology on which the theory is based. After all, what does the ontology look like given inhomogeneous initial conditions? Getting clear on exactly how the model violates SI in the preferred sense is crucial, as a key claim is that violations of SI are restricted to quantum experiments, and do not arise, for example, in randomized clinical trials.\footnote{See \textcite[p460]{ciepielewski2023}. See also \textcite[sec. 6]{hossenfelder2020superdeterminism}, who appeals to LQM as a solution to this problem.} This point is crucial to whether superdeterminism can be considered a truly scientific theory (discussed in section \ref{scientific}). We do not think this is impossible to show, but think more work needs to be done to make this clear: given this ontology, what exactly does a distribution of $\lambda$s and corresponding settings $a$ and $b$ look like for trials of a Bell experiment, so that we can see an explicit correlation between them? Either way, we agree that LQM provides a useful framework to advance the debate over superdeterministic violations of statistical independence.

\subsubsection{Statistical flukes}\label{flukes}

This section considers a class of superdeterministic theories that do not involve fine-tuned initial conditions, but instead violate SI through statistical flukes. There are both deterministic and indeterministic examples.

To understand how this is possible, we first must be clear on what is meant by ``theory". In the previous section, we discussed LQM. However, LQM given inhomogeneous initial conditions does not violate SI. What violates SI is LQM plus an additional postulate: homogeneous initial conditions. So strictly speaking, the ``theory" that violates SI is really a physical theory (in the ordinary sense, i.e., dynamics without boundary conditions) plus a postulate. The same idea will apply here in this section, except instead of an additional postulate about initial conditions, we have additional postulates about flukes. A superdeterministic theory based on flukes is therefore to be understood as a conjunction of a physical theory plus a fluke postulate. A fluke postulate does not concern an initial boundary condition on the events in spacetime, but the existence of an improbable distribution of events throughout spacetime. 

To begin with, we show how a fluke postulate could violate SI (given the frequency interpretation). We will then move to superdeteministic theories. A simple indeterministic example involves the familiar case of flipping a fair coin.  If we flip one million coins, it is overwhelmingly probable that the number of heads and tails outcomes is roughly equal, which gives a representative sample of the probability distribution we associate with flipping fair coins.  But it is also possible to get all heads, or all tails, or other highly improbable fluke cases that are not representative of the underlying 50/50 probability distribution that was used to randomly generate the sample. If one flips a large number of fair coins and obtains such a highly improbable result, then random sampling has failed to yield a representative sample.

This is an important example, because we can see the violation of SI in two different ways.  First, in the ensemble sense, when we look at typical data from a billion or more actual coin flips, and see that our million heads really do fail to represent the roughly 50/50 distribution in the larger ensemble, and second, without reference to a larger ensemble, we see that the million heads fail to represent the underlying 50/50 probability distribution used to generate the individual outcomes.  In the second case, we have no empirical reference that tells us the distribution should be 50/50, but if we are proposing a particular physical theory with a particular distribution (e.g., 50/50 for fair coin flips), we can see if the ensemble is representative of the proposed distribution. It is also possible that all coins ever flipped will be heads (a ubiquitous fluke), meaning there is no larger ensemble that the million heads fail to represent, so for this case, we can only see the violation of SI in the second way.

It could likewise be that the fact we always see entanglement correlations obeyed in a quantum experiment is a statistical fluke in an indeterministic theory where the correlations are only obeyed with some probability.  Consider such a theory where every time we prepare an entangled state and measure it, the quantum correlations are obeyed for half of all cases, and violated for the other half.  Then seeing the entanglement correlations obeyed a million times in a row is a fluke in exactly the same way as seeing the million heads.  That is, our experimental data were not representative of the true underlying theory (i.e., SI is violated), and have misled us into deriving quantum theory.

As a final indeterministic example, take any deterministic theory that is superdeterministic because of a fine-tuned initial condition, but change it so that the initial condition is selected at random from a distribution for which this fine-tuned initial condition is atypical. In that case, the fact that we see entanglement correlations obeyed in Bell experiments would be a statistical fluke, rather than a fine-tuning, even though the initial conditions and the resulting universe are the same. Note that for some indeterministic theories, we can obtain a deterministic counterpart theory by reinterpreting the indeterministic events in the former as boundary conditions in the latter. In that case, a fluke in the indeterministic theory becomes a fine-tuned boundary condition in the deterministic counterpart.

Deterministic branching theories also produce statistical flukes. A simple example is based on the Everett interpretation (as commonly understood e.g. as in \textcite{wallace2012emergent}), which implies that we have arrived at quantum mechanics in part because we live in a branch whose measurement results confirm the Born rule. There are many other branches in which observers see different results, and so arrive at a different (false) physical theory, or no theory at all (\cite{hemmo2007quantum}). To simplify, if the coins in the above example were quantum coins with 50/50 Born rule probability to be heads or tails, then there is a world with all heads, and another with all tails, even though there are far more worlds with approximately equal numbers.  If an observer finds themselves in a world with all heads, then SI is violated for the same reasons as above, for that observer.  The million heads is a fluke because it fails to represent the 50/50 Born rule distribution that was used to generate all of the $2^{10^6}$ worlds. It would also fail to represent a billion quantum coin flips in the vast majority of worlds, where the number of heads and tails is nearly equal.  And there is also one world where the fluke is ubiquitous, and all quantum coins ever flipped come up heads.  In this world, there is no larger ensemble with roughly 50/50 outcomes that we can say this million coins fail to represent.  All we can say is they fail to represent the underlying Born rule probability in the physical theory.  In this theory, observers in the non-fluke branches are likely to derive quantum theory, while observers in the fluke branches will derive some other theory (e.g., a theory where all quantum coins come up heads).

Finally, there could be a deterministic branching version of the above theory where entanglement correlation rules are not generally obeyed, but can be obeyed for atypical branches.  In this theory, entanglement correlations are obeyed in the atypical fluke branches, and so observers in these branches are likely to derive quantum theory, while observers in the non-fluke branches will derive some other theory (e.g., a theory where entanglement correlations are not obeyed).  Note that atypical fluke branches will occur even for typical initial conditions, so these cases are unrelated to fine-tuning.

These approaches may seem so implausible that they should be left out of the definition of superdeterminism, but we see no strong reason to think that these theories are any less plausible than deterministic theories with fine-tuned initial conditions. Even \textcite{ciepielewski2023} say of their own model that it is not a serious competitor
to more standard interpretations of quantum mechanics. And \textcite{chen2021bell} concludes that it falls short of being an empirically adequate theory that is overall simpler and more attractive than well-known non-local interpretations. We will return to the issue of the tenability of superdeterministic theories in section \ref{consequences}. For now, we turn to our final class of superdeterministic theories.

\subsubsection{Nomic exclusion}\label{Goblins}

In Maudlin's brief discussion of superdeterminism (\cite{maudlin2014bell}), he mentions an alternative to ``massive coincidence" superdeterminism, which all of the above-discussed cases could be seen as examples of. The alternative instead appeals to our ``being manipulated": somehow we are led to choose the settings we choose, or the random number generators we offload this task to are rigged, or some such. Such hypotheses need not involve fine-tuned initial conditions, nor determinism, nor flukes, as we shall show. They may appear highly conspiratorial. But they may also be no worse off in this regard than massive coincidence approaches, and some existing models appear to fall into this category. 

Our third class of superdeterminism violates SI by making certain combinations of measurement settings and ontic states nomologically impossible. We call these ``nomic exclusion models". They allow for the observed data to be explained by a local hidden variable model.  Any complete assignment of local hidden variables to the observables of an entangled state will violate quantum predictions for some measurement settings. To prevent this, these models make it physically impossible to measure those settings, so that quantum predictions are obeyed, but no random selection process will ever obtain a representative sample of the true local hidden variable distribution. Some models in this class simply assign no values to observables that are impossible to measure.

Before we consider existing superdeterministic models that take this form, we consider some fanciful toy models, that abstract away the complex details, and so help us to see exactly how nomic exclusion models violate SI. Our toy models appeal to \textit{goblins}, which are somewhat akin to Maxwell's Demon, who actively manipulate our measurement settings, and/or the hidden variables $\lambda$ that determine the outcomes, in such a way that they are dependent - even if there is no empirical evidence of this manipulation. The goblins are effectively placeholders for the mechanisms of more sophisticated nomic exclusion models. 

The goblins manipulate nature in much the same way that a stage magician deceives their audience.  For example, a stage magician might ensure you choose a particular card from a deck by some trick, even though you think you are free to choose any card.  The key difference is that it is possible with extremely careful observation to see how a stage magician's trick is done, but there is no empirical evidence of the goblin's tricks.  A stage magician's tricks can succeed with no fine-tuned initial conditions of the universe, nor statistical flukes, and the goblin's tricks are no different in this regard.

The goblin theories obey local causality and do not involve retrocausality, because the goblins perform their nefarious deeds at subluminal speeds moving forward in time. In general, the goblins must begin their activities at a past event where they can act as a local common cause for the entire experiment, including the choices of measurement settings.  In this way, the goblins use local hidden variables to produce correlations consistent with the quantum predictions for entangled states, including Bell inequality violations.

Nomic exclusion theories can themselves be categorized into three different forms, depending on the causal order in which the SI-violating correlations are set up. This may not be obvious when looking at existing sophisticated models, but our simple Goblin models help make these structural differences apparent. The Goblin models involve hidden variables ($\lambda$), measurement settings (\textit{X}) of multiple experimenters, and goblins (\textit{G}). They differ in terms of what causes what, where causal relations between $\lambda$, \textit{X}, and \textit{G} are represented below by arrows. We consider three different roles for the goblins, corresponding to three different types of superdeterministic theories.  In all three cases, SI is violated because, for any experiment, not all measurement settings are assigned outcomes consistent with quantum predictions, but the goblins ensure no inconsistent settings can be chosen, thereby preventing the measurement of representative samples that would reveal the discrepancy. 

In what follows we refer to outcomes that correspond to our empirical observations (i.e., outcomes consistent with quantum predictions), as \textit{empirically plausible} outcomes.  
Our goblin theories must assign empirically \textit{implausible} outcomes to \textit{some} measurement settings, because any set of local hidden variables for all measurement settings must violate some of the entanglement correlations.

\textit{Theory 1: $\lambda$$\longrightarrow$G$\longrightarrow$X}. 

Goblins examine the system's hidden variables to determine which settings have empirically plausible outcomes and which do not. They then tamper with the experimenter's brains (or their random number generators, etc.) to ensure that only settings with empirically plausible outcomes are possible. This is a clear case where measurement settings depend on (what determines) the measurement outcomes.

\textit{Theory 2: $\lambda$$\longleftarrow$G$\longrightarrow$X}. 

Goblins choose the system's hidden variables and create empirically plausible outcomes for some settings but not for others.  They then tamper with the experimenter's brains (etc.) to ensure that only settings for which they created empirically plausible outcomes are possible. Here \textit{G} is a local common cause of $X$ and $\lambda$. 

\textit{Theory 3: X$\longrightarrow$G$\longrightarrow$$\lambda$}. 

Goblins first examine physical systems that fully determine which future settings the experimenters choose. The goblins then create an empirically plausible outcome for only that joint setting.  This can be generalized to the case where the future settings are restricted, but not fully fixed, in such a way that a local hidden variable theory can assign empirically plausible outcomes to all of the allowed settings.\footnote{It has been claimed that only a retrocausal theory can depend on the settings in this way (\textcite{wharton2020colloquium, hance2024counterfactual}). However, it is in principle possible for goblins to use information from the past to predict what settings will be chosen with certainty, so that the Goblins can then choose hidden variables that give empirically plausible outcomes for these settings.}

The goblin theories are empirically consistent because the empirically implausible outcomes are systematically hidden from the experimenters.  This means the experimenters never get representative samples of the outcomes assigned to all settings, which violates SI.  The goblin theories also help to illustrate that neither determinism with fine-tuned initial conditions, nor flukes, are necessary for the violation of SI. If the universe - goblins included - is a fully deterministic system, we see no reason to think that the initial conditions must be fine-tuned or atypical for the goblins to act this way.  Besides, our goblins need not be deterministic, and might be acting this way due to their own libertarian free will.

As with superdeterminism based on statistical flukes, superdeterminism based on goblins (or some less fanciful but functionally equivalent mechanism) may seem too implausible, and so not deserving of the label superdeterminism. But again we appeal to the \textit{tu quoque} response: it is far from obvious that fine-tuned initial conditions are any better off. In fact it could be noted that some authors have taken seriously the possibility that we are in a simulation (\cite{bostrom2003we}, \cite{Chalmers2022-CHARVW}); our simulators may act like goblins, to prevent us from discovering the underlying code of the simulation. 

We now move on from the simple goblin toy models to more sophisticated models that have been proposed in the literature, which fall under the nomic exclusion category. The models of \textcite{donadi2020superdeterministic} and \textcite{hance2022wave} are consistent with goblin theory 2 ($\lambda$$\longleftarrow$G$\longrightarrow$X) and goblin theory 3 (X$\longrightarrow$G$\longrightarrow$$\lambda$). For in these models, the physical state $\lambda$ at the source depends explicitly on the future setting choice $X$, and this state does not specify outcomes for any other measurement settings.  Therefore, no other choices of the measurement settings are physically possible. They do not specify how this dependence comes about in their model. In theory 2, using the goblin visualization for simplicity, a goblin chooses and fixes the future setting $X$ well in advance, and then prepares the ontic state $\lambda$ so that entanglement correlations will be obeyed for setting $X$.  Outcomes for other measurement settings are not even defined for this $\lambda$.  The goblin still produces the standard quantum long-run statistics for random choices of measurement settings and their outcomes, but SI is clearly violated because the cases that are measured by each different setting $X$ have distinctly tailored $\lambda$ (i.e., $\rho(\lambda|X) \neq \rho(\lambda|X')$).  

The counterfactually restricted theory of \textcite{hance2024counterfactual}, which is closely related to \textcite{hance2022supermeasured}, is a nomic exclusion model that is consistent with goblin theory 1, since observables with empirically implausible outcomes are nomologically impossible to measure, while it is possible to measure any observables with plausible outcomes. We discuss the (2022) model in more detail in section \ref{Palmer}.

Finally, \textcite{ciepielewski2023} consider an alternative version of LQM, where the superdeterminism does not arise from initial conditions or flukes, but instead from the laws. In particular, they start from generic initial conditions and then show that, from the dynamics alone, one can arrive at the condition of homogeneity that correlates settings and $\lambda$.  Another way to say this is that the homogeneous state is an attractor of the theory, which it will settle into for a wide variety of typical initial conditions.  This can be seen as a version of Goblin theory 2, where the Goblins (here, the laws) act over time to create superdeterministic correlations.

\subsection{Do retrocausal and invariant set theories violate SI?}\label{mistakenidentity}

\subsubsection{Retrocausal theories}\label{retrocausal}

It is often said that there are two quite distinct ways of maintaining locality by violating SI, one is superdeterminism, the other appeals to \textit{retrocausality}. Since we have defined all systematic violations of SI to imply superdeterminism, this would mean that retrocausality is also a type of superdeterminism.\footnote{\textcite{sep-bell-theorem} and \textcite{ciepielewski2023} are examples of the standard view, according to which retrocausality violates SI without being superdeterministic. \textcite{nikolaev2023aspects} is an exception, as they treat retrocausality as superdeterministic.} However, in what follows, we argue that retrocausality, properly understood, does not violate SI, and it is for this reason that retrocausality is not superdeterministic.

We will argue that there are two different ways of interpreting retrocausal models, in terms of \textit{causal order} or  \textit{temporal order} and that while SI does appear violated in the more standard temporal order interpretation, the causal order interpretation is preferable, and need not violate SI. The $\lambda$s in $\rho(\lambda) = \rho(\lambda|Z)$ are supposed to represent the relevant ontic states provided by the physical theory. A key premise in our argument will be that the ontic states relevant to SI will generally differ between the causal and temporal order interpretations.
We should also stress that the ontic states $\lambda$ in retrocausal theories do not obey Bell's notion of local causality, which are predicated on causal influences obeying temporal order and never connecting spacelike separated events.

As a simple example of a retrocausal toy model, consider the Zig-Zag model of \textcite{costa1976time}.\footnote{For other retrocausal models, see \textcite{sutherland1983bell}, \textcite{sutherland2008causally}, \textcite{price2015disentangling}, \textcite{wharton2018new}, and \textcite{wharton2020colloquium}.} In this model of a Bell experiment, the causal chain begins at the source, goes to Alice's measurement, goes back (in time) to the source, and then finally goes to Bob's measurement, with none of these links directly connecting space-like separated events.  This model explains the results of the Bell experiment using local retrocausality, and the entanglement correlation is produced because Alice's result is sent back to the source where the other qubit is then prepared in the properly correlated state before being sent to Bob.

What follows is a formalization of this model using standard Born rule probabilities, but for local ontic states. Given the prepared Bell state $|\Psi\rangle_{12} = \frac{1}{\sqrt{2}}\big(|0\rangle_1 |1\rangle_2 - |1\rangle_1 |0\rangle_2\big)$, the reduced density matrix of the first qubit is the maximally mixed state $\hat{\rho}_1$.  This is defined to be the local ontic state of Alice's qubit in this model, i.e., $\hat{\rho}_1 = \lambda_1$. Alice chooses a measurement setting $X$ and measures her qubit, obtaining an outcome with standard Born rule probability given $\hat{\rho}_1$.  Alice's outcome state $|k\rangle_1$ is sent retrocausally back to the source event, and the other qubit is then prepared in the properly correlated state obtained by projecting the Bell state onto Alice's outcome and renormalizing, $|\phi\rangle_2 = \sqrt{2}\langle k|_1 |\Psi\rangle_{12}$.  That qubit is then sent to Bob, and this quantum state is defined to be the local ontic state $|\phi\rangle_2 = \lambda_{2}$ of that qubit.  Bob chooses a measurement setting $Y$ and measures his qubit, obtaining an outcome with standard Born rule probability given $|\phi\rangle_2$.  Bob's outcome obeys the entanglement correlations because of how $|\phi\rangle_2$ is defined from the Bell state using Alice's outcome.

We now show how SI is violated by our retrocausal Zig-Zag under the standard temporal order interpretation.  In this picture, both $\lambda_1$ and $\lambda_2$ exist at the source, so the complete ontic state at the source is $\lambda  = \lambda_1 \cup \lambda_2$.  The values of $\lambda_2$ are perfectly correlated with Alice's measurement setting $X$, because they are created using the outcome of Alice's measurement.  As a result, the equation $\rho(\lambda_2|X) = \rho(\lambda_2)$ is violated, and thus by extension the SI equation $\rho(\lambda|X) = \rho(\lambda)$ is also violated.  In other words, if Alice randomly separates her ensemble into two sub-ensembles, where all elements in one bin will be measured by one setting, and all elements in the other by another setting, the $\lambda_2$s in each sub-ensembles would all match (predetermine) her setting, and the two sub-ensembles would not be representative of the overall distribution of $\lambda_2$ values in the ensemble.  

We now show how SI is \textit{not} violated under the causal order interpretation. At the first step in the causal order of this model, the state $\lambda_1$ has been prepared, but $\lambda_2$ does not yet exist, because Alice has not yet chosen her setting or made her measurement, so the complete ontic state at this step is $\lambda = \lambda_1$.   According to this model, when Alice determines her choice of setting, she always gets representative samples of $\lambda_1$ for any setting, because $\rho(\lambda_1|X) = \rho(\lambda_1) = 1$, and it follows that the SI expression $\rho(\lambda|X) = \rho(\lambda)$ is obeyed.  After Alice performs her measurement with setting $X$, and $\lambda_2$ is created using the outcome, $\rho(\lambda_2|X) = \rho(\lambda_2)$ is still violated as before, but this is just the expected correlation between a setting and an outcome \textit{after} a measurement has been made, so does not actually indicate a violation of SI.  In temporal order, $\lambda_2$ exists before $X$ is chosen, but in the Zig-Zag causal order, $\lambda_2$ only exists after Alice's measurement has been performed.  A simple way to see why violation of $\rho(\lambda_2|X) = \rho(\lambda_2)$ does not constitute an SI violation is that $X$ is no longer a random selection after the measurement.  By using selection $X$, one knows the $\lambda_2$ in each sub-ensemble will be eigenstates of a particular setting observable.

In the final step, Bob chooses his setting and performs his measurement.  Even though $\lambda_2$ may take on different values from run to run, it is still the case that $\rho(\lambda_2|Y) = \rho(\lambda_2)$ because nothing depends retrocausally on Bob's randomly selected measurement setting $Y$.  As a result, Bob succeeds in obtaining a representative sample of the overall distribution of $\lambda  = \lambda_1 \cup \lambda_2$, and thus SI is also obeyed for Bob's measurement, i.e., $\rho(\lambda|Y) = \rho(\lambda)$.

Thus, we have shown that when SI is evaluated following the causal order of a retrocausal theory, it is not necessary violated (although a retrocausal theory could certainly be superdeterministic for other reasons, such as a fluke postulate).  We conclude that because SI depends explicitly on the ontic states $\lambda$ provided by a theory, it makes more sense to evaluate SI in the causal order in which $\lambda$ evolves within that theory. This applies to all physical theories, not just this Zig-Zag model. We think that if we apply it to any theory, retrocausal or not, and find that SI is violated, it is correct to conclude that the theory is superdeterministic.

Proponents of retrocausal models have argued that generalized relativistic locality can be preserved, even allowing some causal influences to move backward in time between lightlike or timelike separated events, provided none ever directly connect spacelike separated events.  This allows indirect causal chains to connect spacelike separated events, as in this example.  The Zig-Zag model above obeys this generalized locality principle, as well as SI, so a reformulated version of Bell's theorem using this locality principle would need a separate assumption to rule out retrocausal models.  This is analogous to how Bell did not initially acknowledge SI as a necessary assumption, until it became clear that violating SI allows for local hidden variable theories.\footnote{This is also analogous to the eventual acknowledgment that a ``one world" assumption is necessary too, see footnote 1.}

\subsubsection{Invariant set theory}\label{Palmer}

The ``invariant set theory" model is defended in \textcite{hance2022supermeasured} and in \textcite{palmer2024superdeterminism}. The model is intended to be a local, superdeterministic interpretation of quantum mechanics. However, while we agree that the model presented in (\citeyear{hance2022supermeasured}) is local and superdeterministic, we find that the modifications proposed in (\citeyear{palmer2024superdeterminism}) result in a model that is neither local nor superdeterministic. 

In the (\citeyear{hance2022supermeasured}) paper, the authors make clear that the existence of a specific invariant set $\lambda$ makes it nomologically impossible for Alice and Bob to choose certain pairs of settings, which clearly leads to a violation of SI (see goblin theory 1 in section \ref{Goblins}).  The impossibility of those settings allows for an entirely local hidden variable model to explain the empirical data in a Bell experiment.  In this version, the invariant set is determined before Alice and Bob make their choices.

In the more recent paper, \textcite{palmer2024superdeterminism} insists that Alice and Bob are free to choose any setting, and that their choices are part of the all-at-once (across all of space and time) selection of a compatible invariant set, so their choices are \textit{ontologically prior} to the selection of the invariant set - i.e., the choices are made `before' the invariant set is determined. Palmer thus adopts an ``all-at-once" model (\cite{adlam2022two}), where no ``initial" condition or state is privileged and gives rise to later states. All states of the universe instead come together at once. This process involves synthesizing information from many spacelike separated events, and then distributing correlated outcomes to spacelike separated events, and is thus highly nonlocal.

In Palmer's model, given choices $X$ and $Y$, certain values of $\lambda$ cannot be selected by nature.  In other words, the order of ontological priority has been reversed, and the free choices $X$ and $Y$ are determined `before' the invariant set $\lambda$.\footnote{For more on ``all-at-once" freedom, see \textcite[Sec. 5.2]{waegell2023generative}.}

Palmer's argument that his theory violates SI assumes that $\rho$ is an ontic probability. Since his theory is deterministic, the ontic probabilities are all zero or one. His argument for the violation of SI can be concisely reconstructed as follows (where the measurement settings take binary values 0 and 1, and $Y'= 1-Y$):

\begin{quote}
(1) $\rho(\lambda|X,Y) = 1$. [From the theory's determinism.]

(2) $\rho(\lambda|X,Y') = 0$. [From the theory's ``counterfactual indefiniteness".]

(3) If SI then $\rho(\lambda|X,Y) = \rho(\lambda|X,Y')$.  

(4) So, SI is false.
\end{quote}

Premise (3) holds because SI requires $\rho(\lambda) = \rho(\lambda|Z)$ for any apparently random selection $Z$. 

Premises (1) and (2) come from invariant set theory. Alice and Bob measured $X$ and $Y$ which are the settings corresponding to ontic state $\lambda$. Had Bob measured $Y'$ instead, the ontic state $\lambda$ would have been different, because it is nomologically impossible to measure $Y'$ given the original $\lambda$.  Thus, if Alice and Bob are truly free to measure any settings, then a compatible value of $\lambda$ must be determined nonlocally using both of their spacelike separated settings, such that the entanglement correlations are obeyed by the eigenvalues in $\lambda$.  The eigenvalues of the measured settings are then nonlocally distributed to the spacelike separated measurement outcome events.  Essentially, by selecting their settings $X$ and $Y$, Alice and Bob obtain a correlated value of $\lambda$ as an outcome.

Let us now explain how the all-at-once selection process is supposed to work in the model.  First, to simplify, imagine that free settings $X$ and $Y$ are the only measurements ever performed in the universe. Then according to \textcite{palmer2024superdeterminism}, reality takes a certain form $\lambda$, which determines eigenvalues for these measurement settings that conform to quantum mechanics. This gives premise (1). This particular $\lambda$ will not (according to this theory) give eigenvalues for counterfactual setting $Y'$, which gives premise (2). We now drop the simplification (of there being only two measurements) and return to Palmer's full theory, allowing many measurements across the universe. Then his idea is that all such measurements determine a particular ontic state $\lambda$ all at once.  As before, this $\lambda$ will not give outcomes for counterfactual settings, which again leads to premise (2).

While we agree that the ontic state $\lambda$ is correlated with the setting choices $X$ and $Y$, this appears to be because $\lambda$ is essentially the measurement outcome, and it is natural to expect measurement outcomes to be correlated with measurement settings. To address SI, we need to look at the process by which a particular invariant set $\lambda$ is determined using Alice's and Bob's free choices, and the ontic state $\lambda_0$ that existed `before' $\lambda$.  Palmer does not introduce $\lambda_0$, but it is implicit in his model. This is because if Alice and Bob are genuinely free, then they could have measured otherwise, in which case, there would have to be facts about what outcomes they would get, had they chosen these alternative settings. Something ($\lambda_0$) must determine these counterfactual outcomes. As such, it is not clear that SI is violated by these measurements, with respect to the ontic state $\lambda_0$, which exists prior to any measurements being chosen.  Palmer is clear that Alice and Bob are completely free, meaning that the choice of measurement settings is not restricted by conditioning on $\lambda_0$, i.e., $p(X,Y) = p(X,Y|\lambda_0)$.  

What we think Palmer's argument in (1)-(4) really shows is not a violation of SI, but a violation of a principle that is sometimes called \textit{counterfactual realism}, where eigenvalues exist for all observables, even those that are not measured.  This principle conflicts with quantum uncertainty principles, quantum contextuality \textcite{KS}, and most directly the Leggett-Garg inequality (\cite{leggett1985quantum}), so its violation is not particularly surprising.  On the upside, this means that Palmer's 2024 theory, while nonlocal, is perfectly consistent with the scientific method, which may be a problem for genuinely superdeterministic theories (see section \ref{scientific}).

\section{Philosophical consequences of superdeterminism}\label{consequences}

Having spelled out a broader space of possible superdeterministic theories, we now look to philosophical consequences, and try to resolve some disputes in the literature. In particular, we focus on the issues of free will (\ref{freewill}), conspiracy (\ref{conspiracy}), and scientific method (\ref{scientific}). Each section is self-contained and can be read independently of the others.

\subsection{Is superdeterminism compatible with free will?}\label{freewill}

Bell often described superdeterminsim as a threat to free will. For example, in his final paper on his theorem (\cite{bell1990nouvelle}), he says,

\begin{quote}
    ``An essential element in the reasoning here is that [the measurement settings] are free variables. One can envisage then theories in which there just are no free variables [...] In such ‘superdeterministic’ theories the apparent free will of experimenters, and any other apparent randomness, would be illusory.''
 \end{quote}

Bell has since been frequently accused of ``causing confusion" with such remarks (\cite{baas2023does}, \cite{allori2024hidden}), and it now seems widely accepted that superdeterminism has nothing to do with free will (see also \cite{esfeld2015bell}). The following argument, from \textcite{baas2023does}, represents a commonly held view:

\begin{quote}
``superdeterminism is the conjunction of determinism and the atypicality of cosmological initial conditions [...] and as such, is no more problematic for free will than any deterministic theory. If one accepts the incompatibilist claim that free will is incompatible with determinism, it follows trivially that free will is incompatible with superdeterminism. In fact, there are many ways to reconcile free will with determinism, namely to endorse a form of compatibilism (for an overview, see e.g. McKenna and Coates [2021]). If free will is compatible with determinism, it is not clear why it should be incompatible with superdeterminism."
\end{quote}

The idea is that if superdeterminists simply embrace compatibilism, the view that free will is compatible with determinism, then they will be no worse off than ordinary determinists. Here, we argue that this is insufficient. For there are several prominent versions of compatibilism (described in \cite{sep-compatibilism}) that prevent experimenters in superdeterministic worlds from having the same sorts of freedoms as experimenters in ordinary deterministic worlds. 

According to \textit{classical compatibilism}, Alice's choice to measure $X$ was free if she could have chosen otherwise (e.g. by measuring $Y$). Some classical compatibilists tried to make this idea precise in conditional terms (\cite{Hume1975}, \cite{Ayer1954}, \cite{Hobart1934}). Thus, to say that Alice could have measured $Y$ and not $X$ is to say that, had she wanted (chosen, willed, or decided) to measure $Y$ and not $X$ at that time, then she would have measured $Y$ and not $X$. In a merely deterministic theory, Alice's alternative want is physically possible and would lead to that want being satisfied (given the different initial conditions, i.e., the different wants). 

But in a superdeterministic theory, this may not be the case, especially in the context of nomic exclusion models. Consider the example described in \textcite{chen2021bell}, in which entangled photon pairs are either absorbed or transmitted by polarizers that are oriented in various different directions. Assume the left polarizer is set at 0 degrees, while the right polarizer is set at 30 degrees. Following \textcite{andreoletti2022superdeterminism}, call this ``set-up A", and call the actual photon ensemble measured by this set-up ``sub-collection \textit{a}". According to \textcite{andreoletti2022superdeterminism}, superdeterminism entails that if set-up B (where both polarizers are rotated another 30 degrees each) had instead been chosen, then a completely different photon ensemble (call it $b$) would have been measured. This is clear in goblin theory 3 (X$\longrightarrow$G$\longrightarrow$$\lambda$) from section \ref{Goblins}, where the goblins select ensembles to be measured based on what measurement settings will be chosen. Now take the experimenter who chooses set-up A, and who therefore measures ensemble \textit{a}. If the experimenter believes she could have measured \textit{this} ensemble (sub-collection $a$) with set-up B, then she is just wrong. She is not free to do otherwise. So her choice of set-up A for sub-collection $a$ was not free. Similar reasoning applies to the other two goblin theories. Similar reasoning also applies to fine-tuned versions of superdeterminism, especially if the postulate about the initial conditions is a genuine law or constraint (\cite{baas2023does}). 

Note that the limitation on free will is restricted. The experimenter could not have chosen set-up B \textit{for sub-collection $a$}, meaning she was not free according to classical compatibilism.  But there are plenty of other things she is free to do. But this is a clear restriction on the kind of freedom we think we have, and it is a freedom that classical compatibilism will deliver for ordinary deterministic theories, where set-ups and sub-collections are independent. Superdeterminism - at least in the context of nomic exclusion models - therefore restricts our freedom more than ordinary determinism does, according to classical compatibilism. 

Classical compatibilism has fallen out of favour among modern compatibilists, largely due to some apparent counterexamples to the conditional analysis (\cite{Chisholm1964}, \cite{vanInwagen1983}). If I have a crippling fear of snakes, then I cannot freely pick up a snake, even though, if I had wanted to pick one up, I would. Nonetheless, some modern versions of compatibilism still return a result that would vindicate Bell's concerns about free will. An example is what \textcite{sep-compatibilism} call \textit{new dispositionalism}, advocated for example by Vihvelin (\citeyear{vihvelin2004free}, \citeyear{vihvelin2013causes}) and \textcite{smith2003rational}. 

According to the new dispositionalists, we hold fixed the relevant causal base or underlying structure of an agent’s disposition to do something (like pick up a snake or implement set-up B on sub-collection $a$), and we consider various counterfactual conditions in which that causal base or underlying structure operates unimpaired. Does the agent in an appropriately rich range of such counterfactual conditions implement different set-ups on the same ensemble, or handle snakes? If so, then even if in the actual world she does not implement set-up B (but used A instead), or does not pick up a snake, she was able to do so. For she had at the time of action the pertinent agential abilities or capacities, even if the world is determined. But if a nomic exclusion version of superdeterminism is true, then the relevant counterfactual conditions do not allow one's disposition to implement set-up B on ensemble \textit{a} to ever manifest. Thus, modern versions of compatibilism that are sensitive to the kinds of alternative possibilities that prominent superdeterministic models rule out, will support Bell's concerns about free will.

It should be noted that there are also many modern compatibilist views which disassociate free will from the ability to perform alternative actions completely. For example, \textcite{Frankfurt1971}'s \textit{hierarchical mesh theory}, which explains freely willed action in terms of actions that issue from desires that mesh with hierarchically ordered elements in one's psychology. On this view, free actions stem from desires nested within more encompassing elements of oneself,
such as second order desires, desires to have certain desires, or desires to be a certain sort of person. If my implementation of set-up A, came from desires within me, and about the type of rigorous experimenter I want to be, then I freely implemented that experiment, whether or not there are any goblins, or fine-tuned initial conditions. We therefore do not intend to claim that superdeterminism undermines free will. We only claim that it is an open question, which is not resolved merely by pointing to compatibilism. 

To conclude, we consider a quite different argument for the independence of superdeterminism and free will from \textcite{allori2024hidden}:

\begin{quote}
 ``More generally, one can see that the issue of free will is of no importance in these matters by observing that to define what SI is one needs no human beings. In fact, while it is the case that usually humans select the experimental settings, these could well be chosen by some sort of automatic random generator. If that is the case, then no genuine choice is involved or needed to define the hypothesis of SI."
 \end{quote}

But this argument is not valid. We can similarly observe that to define what determinism is, one needs no human beings. But that does not stop determinism from being important to the issue of free will.

\subsection{Is superdeterminism conspiratorial?}\label{conspiracy}

A common objection to superdeterminism is that it is \textit{conspiratorial}. Perhaps the first to articulate this objection was \textcite{Shimony1976}, who noted that 

\begin{quote}
``In any scientific experiment in which two or more variables are supposed to be randomly selected, one can always conjecture that some factor in the overlap of the backwards light cones has controlled the presumably random choices. But, we maintain, skepticism of this sort will essentially dismiss all results of scientific experimentation. Unless we proceed under the assumption that hidden conspiracies of this sort do not occur, we have abandoned in advance the whole enterprise of discovering the laws of nature by experimentation."
\end{quote}

Similar points are pressed by \textcite{maudlin2014bell}, \textcite{sen2020superdeterministic2}, \textcite{chen2021bell}, \textcite{allori2024hidden}, and others. Meanwhile, authors sympathetic to superdeterminism have denied that superdeterminism is conspiratorial in any problematic sense. 

 \textcite{hossenfelder2020rethinking} interpret what they call ``the conspiracy argument" as pointing out that the range of initial conditions that result in the relevant correlations between settings and $\lambda$ are narrow. So to respond to it, they consider a theory with a constrained state space and argue that, within such a space, a physically relevant measure might reveal that the set of initial conditions yielding the correlations is broad. We do not find this response compelling. Such a theory will exclude possibilities which will \textit{inevitably seem possible to observers}, such that according to the theory, nature is systematically deceiving us. This, as we will argue, is still conspiratorial.

More recently, \textcite{andreoletti2022superdeterminism} have tried to rebut the conspiracy objection. They argue that the objection can be understood as the claim that superdeterminism entails the truth of ``suspicious counterfactuals". They consider the example described in \textcite{chen2021bell}, in which entangled photon pairs are either absorbed or transmitted by polarizers that are oriented in various different directions. Oriented in one way, quantum mechanics (correctly) predicts a certain fraction of photons will be absorbed. Oriented in another way, quantum mechanics predicts
a different fraction. Assume the left polarizer is set at 0 degrees, while the right polarizer is set at 30 degrees. Andreoletti and Vervoort call this ``set-up A". They refer to the actual photon ensemble measured by this set-up as ``sub-collection \textit{a}". They then note that according to superdeterminism, if set-up B (where both polarizers are rotated another 30 degrees each) had instead been chosen, then a completely different photon ensemble (call it $b$) would have been measured. Thus, superdeterminism apparently entails ``suspicious counterfactuals" such as (C):

\begin{quote}
   (C) If the set-up B had been chosen, then the sub-collection \textit{b} would have been selected. 
\end{quote}

And this seems bizarre! For surely, if we had chosen set-up B instead of A, our ensemble would not have been different, we would still be experimenting on the same ensemble. Thus, Andreoletti and Vervoort argue that

\begin{quote}
    ``the charge of conspiracy against superdeterminism lies \textit{precisely} in counterfactuals such as (C). That is, people who find superdeterminism a conspiracy theory and hence a non-starter do it because they think superdeterminism implies (C) and (C) is just hard to believe."
\end{quote}

Let us grant for the moment that this is what the conspiracy charge amounts to. Then, Andreoletti and Vervoort offer a response that is not unreasonable. They say that (C) is the wrong counterfactual to be focusing on. The correct one to consider is (C*): If the set-up B had been chosen \textit{and nature is local}, then the sub-collection $b$ would have been selected. So if we insist, as superdeterminists typically do, that nature is local, then the  fact that sub-collection $b$ would have been selected should not seem so mysterious.

However, as we have already suggested, this is not the right way to think about the conspiracy objection. To establish that a superdeterministic theory is conspiratorial, we need a clear and intuitive sufficient condition for a theory being ``conspiratorial". We then need to show that the superdeterministic theory satisfies the condition. We propose that a theory is conspiratorial if it entails that experimenters are \textit{systematically deceived} by nature.

Whether it happens by fine-tuned initial conditions, statistical flukes, or nomic exclusion, superdeterminism implies that our experiences will lead us to believe that we are able to perform certain experiments on representative ensembles, when in fact we cannot.\footnote{Note that if these experiences systematically lead us to believe in libertarian free will, so that we ``could have done otherwise'' in a sense that's incompatible with determinism, then any deterministic theory will seem conspiratorial by our definition. But belief in libertarian free will is hardly systematic. Given the prevalence of compatibilism, we cannot therefore say that given determinism, we are \textit{deceived} into believing we have free will, because on the compatibilist view, we actually do have free will. Deterministic theories are therefore not in general conspiratorial, according to our definition.} For example, we are simply unable to use set-up B to measure sub-collection $a$. Yet, every experiment we've performed in the past has led us to believe that such a feat is entirely possible. For we have performed experiments just like $B$ on ensembles that seem just like $a$ in the past. So we are led to believe we can do it, but we never actually can.  We think this is manifestly conspiratorial. Does this mean superdeterminism is unscientific? We will consider this in the next section.

\subsection{Is superdeterminism unscientific?}\label{scientific}

The ``objection from the impossibility of science" (as \textcite{andreoletti2022superdeterminism} call it), has been posed by many authors, in a variety of ways, for example, \textcite{Shimony1976}, \textcite{baas2023does}, \textcite{Goldstein2011}, \textcite{chen2021bell}, and \textcite{allori2024hidden}.  

\textcite{Shimony1976} argue that denying SI is ``wrong on methodological grounds'' if no specific causal linkage is proposed. For unless we ``proceed under the assumption that hidden conspiracies of this sort do not occur, we have abandoned in advance the whole enterprise of discovering the laws of nature by experimentation''. \textcite{maudlin2014bell} also argues that denying SI ``would undercut scientific method.'' The reason is that all ``scientific interpretations of our observations presuppose that they have not have been manipulated in such a way.'' For example, if every mouse that is exposed to cigarette smoke is found to get cancer, then we may inductively infer that cigarette smoke causes cancer in mice generally. But if our mice samples are not representative of mice (e.g. because goblins tamper with the lungs of our samples), then we cannot use induction at all, and science is ruined.  \textcite{baas2023does} go a step further, and argue that if SI were false (generally) in a theory, then it would suffer from the problem of \textit{empirical incoherence}: its truth undermines our reasons to believe it (\cite{barrett1996empirical}). If the theory is true, then nature causes us to believe false theories. 

Somehow, superdeterminists must justify the restriction of SI violation to Bell experiments. \textcite{chen2021bell} concedes that ``it is logically consistent for one to claim that statistical independence is false about microscopic systems but for all practical purposes true of macroscopic systems". He considers decoherence as a possible mechanism for explaining how this could be, but concludes that it cannot offer any such explanation, and so suggests that there is a serious challenge here for explaining why SI violations aren't ubiquitous. 

\textcite{baas2023does} introduce the term \textit{exceptionalist SD} to refer to superdeterministic theories in which SI is obeyed in almost all circumstances, with the only exception being measurement settings for entangled quantum states, like those used in a Bell test.  Most superdetermininistic theories in the literature are intended to be exceptionalist, so that it remains safe to assume SI in all other areas of science.  Because exceptionalist SD theories have cases where SI is obeyed, \textcite{baas2023does} claim that ``they must assert the impossibility of using those systems in order to set the measurement settings [in a Bell test]."  Since it is otherwise unclear how one might `use those systems,' we assume Baas and Le Bihan mean the systems used to perform the random selection procedures for cases that obey SI. An exceptionalist SD theory would then need to make the same random selection procedure impossible for choosing measurement settings in a Bell test.  But we think this is incorrect. For example, we could flip coins and use the results of the same coins to choose settings in a Bell test and also to assign subjects to different groups in a drug trial, and this could result in representative samples for the subjects in the trial, while failing to produce representative samples for the Bell experiment.
That is, in such a theory, we could have $\rho(\lambda_{\textrm{drug}}) = \rho(\lambda_{\textrm{drug}}|X)$ and $\rho(\lambda_{\textrm{Bell}}) \neq \rho(\lambda_{\textrm{Bell}}|X)$, where  $\lambda_{\textrm{drug}}$ are the ontic states that determine how a subject in a drug trial responds to the treatment they receive, and $\lambda_{\textrm{Bell}}$ determine the outcomes of measurements in a Bell experiment.
So exceptionalist SD may be a logically consistent option, but given that the validity of the entire scientific method is at stake, it behooves proponents of exceptionalist SD, like \textcite{andreoletti2022superdeterminism} and \textcite{nikolaev2023aspects}, to provide the kind of ``specific causal linkage'' that \textcite{Shimony1976} saw was lacking, i.e., to explain and justify the exception to the SI rule.

What then are we to make of the scientific status of superdeterminism? Should we deem it unscientific or pseudo-scientific? Here we think it is instructive to consider the widely accepted framework of \textcite{lakatos1970falsification}, which is designed to answer such questions. Lakatos urges us to think less in terms of particular theories, and more in terms of \textit{research programmes}. A research programme is a sequence of falsifiable theories characterized by a shared core of assumptions that the programme takes for granted. For superdeterminism, these assumptions may include limited violations of SI and locality. A research programme then constitutes good science (science that is rational to work on and develop) if it is \textit{progressive}, and bad science, if it is \textit{degenerating}. Progressive programmes must meet two conditions. First, they must be theoretically progressive: each new theory in the sequence must have excess empirical content over its predecessor; it must predict novel and hitherto unexpected facts. Second, they must be empirically progressive. Some of that novel content has to be experimentally confirmed. If these conditions fail, the programme is degenerative.  

The superdeterminism research programme cannot be described as progressive. For although some experimental tests for some superdeterministic models have been proposed (e.g. \cite{Hossenfelder2011-HOSTSH}), they are not compelling. But it is also too early, we think, to consider superdeterminism inevitably degenerative. The mere logical possibility of a limited sort of statistical dependence in certain quantum experiments shows \textit{there is no proof} that superdeterminism entails ubiquitous violations of SI. There is also no proof that we could not one day empirically discover the conspiratorial dependencies (e.g., we uncover the ``goblins''). 

Superdeterminism, presently, is therefore neither scientific nor unscientific, it is better thought of as being in a \textit{pre-scientific} stage, where we are still coming to grips with what such models require, and we are only just beginning to propose simple toy models, several of which have been analyzed here. We therefore see no conclusive reason why the superdeterminism programme should be abandoned, as it may yet yield scientific fruit.

\section{Conclusion}\label{conclusion}

In this paper we have adopted and motivated the frequency interpretation of SI and we have argued that SI-violating superdeterministic models are not confined to deterministic models with atypical or fine-tuned initial conditions. We argued for two additional categories of superdeterminism, one involving fluke postulates, the other involving nomic exclusion, and we showed that several existing models fall into the latter category. We then argued that the attempt to categorize retrocausal models and invariant set theory as SI-violating is more subtle than has been appreciated, and depends upon exactly how we understand these models. Finally, we argued for three philosophical consequences. First, some compatibilist accounts of free will entail that given superdeterminism, experimenters are not entirely free. Second, superdeterminism is conspiratorial in the sense that it postulates that nature is deceiving us. Third, superdeterminism is not unscientific but pre-scientific. We hope this motivates further research into the real meaning of SI-violation and the space of possible superdeterministic theories. \newline\newline

\textbf{Acknowledgments:}  Thanks to Tim Palmer, Wayne Myrvold, and Michael Robinson, for several helpful discussions. This project/publication was made possible through the support of Grant 63209 from the John Templeton Foundation. The opinions expressed in this publication are those of the authors and do not necessarily reflect the views of the John Templeton Foundation.

%\bibliographystyle{apalike}
%\bibliographystyle{IEEEtran} 
%\bibliography{refs}

\printbibliography
\end{document}